\documentclass[12pt]{iopart}
\usepackage{graphicx}
\usepackage{epsfig}
\usepackage{amssymb}
\usepackage{iopams}

\usepackage{dcolumn}% Align table columns on decimal point
\usepackage{booktabs}% For publication quality tznentori-04.pdfables for LaTeX
\usepackage{bm}% bold math
\usepackage{color}

\newcolumntype{d}[1]{D{.}{.}{#1} }
\begin{document}
\title[Toroidal high-spin isomers in light nuclei with $Z$$\neq$$N$]
{Toroidal high-spin isomers in light nuclei with $N$$\neq$$Z$}

\author{A. Staszczak$^1$ and Cheuk-Yin Wong$^2$}

\address{$^1$ Institute of Physics, Maria Curie-Sk{\l}odowska
University, 20-031 Lublin, Poland}
\address{$^2$ Physics Division, Oak Ridge National Laboratory,
Oak Ridge, TN 37831, USA}
\ead{stas@tytan.umcs.lublin.pl \textrm{and} wongc@ornl.gov}

\begin{abstract}
The combined considerations of both the bulk liquid-drop-type behavior
and the quantized aligned rotation with cranked Skyrme-Hartree-Fock
approach revealed previously~\cite{Sta14} that even-even, $N$=$Z$,
toroidal high-spin isomeric states have general occurrences for light
nuclei with 28$\leq$$A$$\leq$52.  We find that in this mass
region there are in addition $N$$\ne$$Z$ toroidal high-spin isomers
when the single-particle shells for neutrons and protons occur at the
same cranked frequency $\hbar \omega$.  Examples of $N$$\ne$$Z$
toroidal high-spin isomers, $^{36}_{16}$S$_{20}$($I$=74$\hbar$) and
$^{40}_{18}$Ar$_{22}$($I$=80,102$\hbar$), are located and
examined. The systematic properties of these $N$$\ne$$Z$ toroidal
high-spin isomers  fall into the same regular
(multi-particle)-(multi-hole) patterns as other $N$=$Z$ toroidal
high-spin isomers.
\end{abstract}


\vspace{2pc}

\submitto{\PS}

%\maketitle


\section{Introduction}
A closed orientable surface has a topological invariant known as the
Euler characteristic $\chi$=$2-2g$, where the genus $g$ is the number
of holes in the surface.  Nuclei as we now know them have the topology
of a sphere with $\chi$=2.  Wheeler suggested that under appropriate
conditions the nuclear fluid may assume a toroidal shape with
$\chi$=0~\cite{Gam61}.  If toroidal nuclei could be made, there would
sprout forth a new family tree for the investigation of the nuclear
species.

In the liquid-drop model, toroidal nuclei are however plagued with
various instabilities~\cite{Won73}, and the search for toroidal nuclei
remains elusive~\cite{Sta08}. When a nucleus is endowed with an angular
momentum along the symmetry axis, $I$=$I_z$, from classical mechanical
point of view, the variation of the rotational energy of the
spinning nucleus can counterbalance the variation of the toroidal
surface energy to lead to toroidal isomeric states at their local
energy minima, when the angular momentum $I$=$I_z$ is beyond a
threshold value~\cite{Won78}.  The rotating liquid-drop nuclei can
also be stable against sausage instabilities (known also as
Plateau-Rayleigh instabilities, in which the torus breaks into smaller
fragments~\cite{Egg97,Pai09}), when the same mass flow is maintained
across the toroidal meridian to lead to high-spin isomers within an
angular momentum window~\cite{Won78}.

The rotating liquid-drop model is useful only as a qualitative guide
to point out the essential balancing forces leading to possible
toroidal figures of equilibrium. Quantitative assessment will rely on
microscopic descriptions that include both the bulk properties of the
nucleus and the single-particle shell effects in self-consistent
mean-field theories, such as the Skyrme-Hartree-Fock (SHF)
approach~\cite{Vau72}.  Self-consistent mean-field theories are needed
because non-collective rotation with an angular momentum about the
symmetry axis is permissible quantum mechanically for an axially
symmetric toroid only by making particle-hole excitations and aligning
the angular momenta of the constituents along the symmetry
axis~\cite{Boh81}.  As a consequence, only a certain discrete,
quantized set of total angular momentum $I$=$I_z$ states are allowed.
The nuclear fluid in the toroidal isomeric state may be so severely
distorted by the change from sphere-like geometry to the toroidal
shape that it may acquire bulk properties of its own, to make it a
distinct type of quantum fluid. The SHF approach is well suited to
describe the changed bulk nuclear properties and their effects on the
stability of the toroidal nuclei.

In our previous work~\cite{Sta14}, we showed by using a cranked SHF
approach that even-even, $N$=$Z$, toroidal high-spin isomeric states
have general occurrences for light nuclei with 28$\leq$$A$$\leq$52.
On the other hand, Ichikawa \textit{et al.}~\cite{Ich12,Ich14a} found
that toroidal high-spin isomer with $I$=60$\hbar$ may be in the local
energy minimum in the excited states of $^{40}$Ca by using a cranked
SHF method starting from the initial ring configuration of 10 alpha
particles. By using different rings of alpha particles,
they subsequently also obtained high-spin toroidal isomers in $^{36}$Ar,
$^{40}$Ca, $^{44}$Ti, $^{48}$Cr, and $^{52}$Fe~\cite{Ich14b},
confirming the general occurrence of high-spin toroidal isomers in
this mass region in Ref.~\cite{Sta14}.

In all these previous studies, the high-spin toroidal isomers are
even-even and $N$=$Z$ nuclei. A natural question arises whether the
high-spin toroidal nuclei are associated with the strong binding of
$\alpha$-particle-type, $N$=$Z$ nuclei.  Are there toroidal high-spin
isomers with $N$$\ne$$Z$? Such a question brings into
focus the related question on the conditions for the occurrence of
toroidal high-spin isomers and how these $N$$\ne$$ Z$ isomers, if
found, fit in the patterns of the systematics of all toroidal
high-spin isomers?  The answers to these questions form the
main subjects of the present investigation.

\section{Shell structure of toroidal nuclei in shell model with radially displaced HO potential}

\begin{figure}[htb]
\begin{center}
\includegraphics[width=0.6\columnwidth]{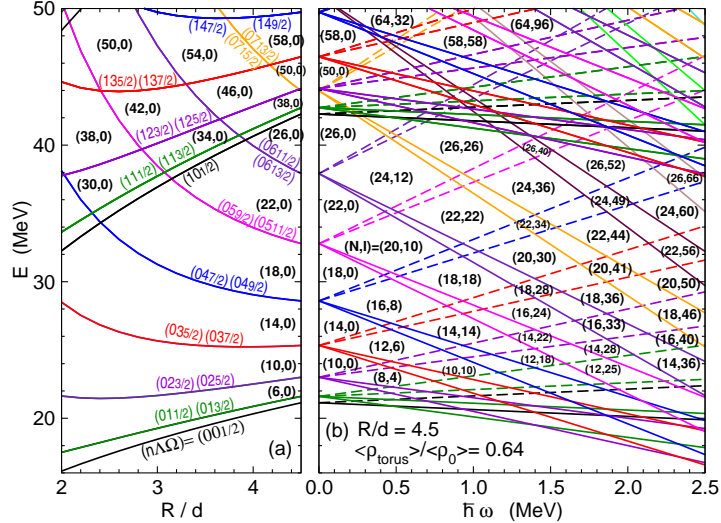}
 \caption{\label{Fig1} (Color online.) (a) Single-particle states of a
   toroidal nucleus with $I$=0 as a function of $R/d$,
   calculated with $\langle \rho_{\rm torus}\rangle /\langle
   \rho_0\rangle$=0.64 and $A$$\sim$40. Each state is labeled by $(n,
   \Lambda, \Omega)$, with $n$=$n_z$+$n_\rho$ and degenerate
   $\Omega_z$.  (b) Single-particle Routhians of a toroidal nucleus
   with $R/d$=4.5, as a function of a cranking frequency $\hbar
   \omega$. Routhians of positive- and negative-$\Omega_z$ states are
   given by the solid and dashed lines, respectively.  The listed pair
   numbers $(N,I)$ refer to the occupation number $N$ and the total
   angular momentum $I$=$I_z$ aligned along the symmetry $z$-axis.
   Figure taken from Ref.~\cite{Sta14}.}
\end{center}
\end{figure}

Our knowledge on toroidal high-spin isomers will be enhanced by
exploring the shell structure of a toroidal nucleus. We need the
single-particle energy diagram in a toroidal nucleus under
non-collective rotation with different aligned angular momenta,
$I$=$I_z$.  For the case of $I$=0, with no rotation, the
single-particle potential for a nucleon in a toroidal nucleus with
azimuthal symmetry in cylindrical coordinates $(r,z)$ can be
represented by the radially displaced harmonic oscillator (HO)
potential~\cite{Won73}
\begin{equation}
V_0(r,z) = \frac{1}{2} m \omega_0^2 (r -R)^2 +\frac{1}{2} m \omega_0^2  z^2,
\end{equation}
where $\hbar\omega_0$=$[(3\pi R/2d)^{1/3}41/A^{1/3}]\langle\rho_{\rm
  torus}\rangle /\langle\rho_0 \rangle$. We have included the ratio
$\langle\rho_{\rm torus}\rangle/\langle\rho_0 \rangle$, where
$\langle\rho_{\rm torus}\rangle$ and $\langle\rho_0\rangle$ are the
average nuclear densities in the toroidal and the spherical
configurations respectively, to take into account the reduced density
in toroidal isomers.
We have neglected the spin-orbit interaction for the low-lying energy
levels as their expectation value of the spin-orbit interaction is
approximately zero~\cite{Won73,Sta14}.
We label a state by $(n\Lambda \Omega\Omega_z)$, where
$n$=$(n_{z}+n_{\rho})$, $\pm\Lambda$ is the $z$-component of the
orbital angular momentum, and $\Omega=|\Lambda \pm 1/2|$ is the
single-particle total angular momentum with $z$-components $
\Omega_z$=$\pm \Omega$.
Fig.~\ref{Fig1}(a) gives the single-particle state energies as a function
of $R/d$ for a toroidal nucleus with $I$=0.

We consider next the single-particle state diagram for a toroidal
nucleus of an aspect ratio $R/d$ under a non-collective rotation with
an aligned angular momentum, $I$=$I_z$. We use a Lagrange multiplier
$\omega$ to describe the constraint $I_{z}$=$\langle \hat{J}_{z}
\rangle$=$\sum_{i=1}^{N} \Omega_{zi}$.  The constrained
single-particle Hamiltonian becomes the Routhian $\hat{h}'={\hat
  h}-\omega \hat{J}_{zi}$, and the aligned angular momentum $I$ is a
step-wise function of the Lagrange multiplier $\omega$ \cite{Ring80},
with each $I$ spanning a small region of $\hbar \omega$. The
single-particle Routhian under the constraint of the non-collective
aligned angular momentum $I$ is
\begin{equation}
E(n \Lambda \Omega \Omega_z)
\sim \hbar \omega_0 (n +1) + \frac{\hbar^2\Lambda^2}{2 m R^2}
-\hbar \omega \Omega_z.
\end{equation}
Fig.~\ref{Fig1}(b) gives the single-particle Routhians as a function
of the constraining Lagrange multiplier $\hbar \omega$, for a toroidal
nucleus with $R/d$=4.5, approximately the aspect ratio for many
toroidal nuclei with 28$\leq$$A$$\leq$52. We can use
Fig.~\ref{Fig1}(b) to determine $I$=$I_z$ as a function of $N$ and
$\hbar \omega$. Specifically, for a given $N$ and $\hbar \omega$, the
aligned $I_z$-component of the total angular momentum $I$ from the $N$
nucleons can be obtained by summing $\Omega_{zi}$ over all states
below the Fermi energy.
The energy scales of the $\hbar \omega$ and $E$ axes in
Fig.~\ref{Fig1}(b) depend on $N$, $R/d$, $\langle \rho_{\rm torus}
\rangle/ \langle \rho_0 \rangle$ which vary individually at different
isomeric toroidal energy minima, but the structure of the $(N,I_{z})$
shells and their relative positions in Fig.~\ref{Fig1}(b) remain
approximately the same in this $A$$\sim$40 mass region. We can use
Figs.~1(a) and \ref{Fig1}(b) as a qualitative guide to explore the
landscape of the energy surface for different $(N,I_{z})$
configurations, by employing a reliable microscopic model.
For a proton/neutron number $N$ such that ($N\bmod{4}$)= 0 or 2,
Table~\ref{table1} gives the simple rules to
calculate an aligned angular momentum, $I$=$I_{z}$, in terms of
particle-hole excitations between the states with $n$=0,
relative to the $I$=0 configuration.

\begin{table}[htb]
\begin{center}
\caption{\label{table1} The quantized values of aligned angular
  momentum $I$=$I_{z}$ for different particle-hole excitations
  between the states with quantum number $n$=0 in the N nucleon
  system.
  The parameter $\Lambda_{max}=[N-(N\bmod{4})]/4$
  is the maximum value of the $z$-component of the orbital
  angular momentum in the ground state.}
\vspace*{0.3cm}
\begin{tabular}{cll}
  \hline
  Excitation & $N$=$4\Lambda_{max}$ & $N$=$4\Lambda_{max}+2$ \\
  \hline
  1p-1h & $I=2\Lambda_{max}  $ & $I=2\Lambda_{max}+2 $ \\
  2p-2h & $I=4\Lambda_{max}+1$ & $I=4\Lambda_{max}+2 $ \\
  3p-3h & $I=6\Lambda_{max}  $ & $I=6\Lambda_{max}+4 $ \\
  4p-4h & $I=8\Lambda_{max}+1$ & $I=8\Lambda_{max}+4 $ \\
  5p-5h & $I=10\Lambda_{max} $ & $I=10\Lambda_{max}+6$ \\
  \hline
\end{tabular}
\end{center}
\end{table}

\section{Toroidal isomers in Skyrme energy density functional approach}

Our objective in the present study is to locate local toroidal figures
of equilibrium, if any, in the multi-dimensional search space of
$(A$=$N$+$Z$,$Q_{20},I)$ with $N$$\ne$$Z$. To do this we used the
Skyrme energy density functional approach where the constraint HFB or
cranked HF equations are solved using the symmetry-unrestricted code
HFODD \cite{hfodd} and an augmented Lagrangian method \cite{alm}.  In
the particle-hole channel the Skyrme SkM* force \cite{Bar82} was
applied and a density-dependent mixed pairing \cite{Dob02,Sta09}
interaction in the particle-particle channel was used in HFB
variant. The code HFODD uses the basis expansion method in a
three-dimensional Cartesian deformed HO
basis. In the present study, we used the basis which consists of 1140
lowest states of deformed HO, having not more than $N_{0}$=26 quanta
in the Cartesian directions.

\subsection{Toroidal configurations with no rotation}

The general occurrence of shell gaps in the single-particle diagrams
in Figs.~\ref{Fig1}(a) and (b) implies that there are many nuclei with
different $N$ and $Z$ that will acquire large shell effects, and these
shell effects provide some degrees of stability for the nucleus.  As a
consequence, we expect from Fig.~\ref{Fig1}(a) that if a nucleus is
constrained to possess a large-magnitude negative quadrupole moment
and the nucleon numbers $N$ and $Z$ reside on the shell gaps of
Fig.~\ref{Fig1}(a), the density of the nucleus will likely be toroidal
in character.

\begin{figure}[htb]
\begin{center}
 \includegraphics[width=0.5\columnwidth]{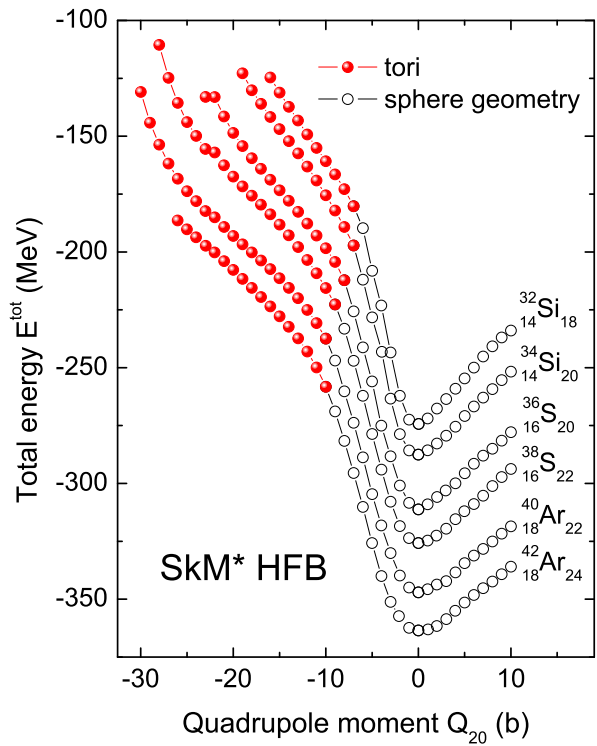}
  \caption{\label{Fig2} (Colour online.) The total HFB energies of
    ${}_{14}^{32}$Si, ${}_{14}^{34}$Si, ${}_{16}^{36}$S,
    ${}_{16}^{38}$S, ${}_{18}^{40}$Ar, ${}_{18}^{42}$Ar, as a function
    of the quadrupole moment $Q_{20}$ for the case of $I$=0.  Toroidal and
    spherical configurations are indicated by full and open circles,
    respectively.}
\end{center}
\end{figure}

Such an expectation is indeed confirmed for the case of $I$=0.  The
Skyrme-HFB calculations for $N$=$Z$ in Fig.~\ref{Fig2} of our earlier
work in~\cite{Sta14}, and for many $N$$\ne$$Z$ nuclei in the present
work in Fig.~\ref{Fig2}, reveal that as the quadrupole moment
constraint, $Q_{20}$, decreases to become more negative, the density
configurations with sphere-like geometry (open circles) turn into
those of an axially-symmetric torus (full circles).  Axially-symmetric
toroidal density distributions are found in ${}_{14}^{32}$Si,
${}_{14}^{34}$Si, ${}_{16}^{36}$S, ${}_{16}^{38}$S, ${}_{18}^{40}$Ar,
and ${}_{18}^{42}$Ar. The presence of the large number of toroidal
configurations when there is a large-magnitude negative quadrupole
moment constraint suggests that the occurrence of these toroidal
configurations are quite common for light nuclei in this mass region,
as predicted by Fig.~\ref{Fig1}(a). As the search is not yet
exhaustive, nuclei with toroidal configurations in addition to those
in Fig.~\ref{Fig2} are possible. Furthermore, because of the
approximate symmetry of $N$ and $Z$, toroidal configurations are also
expected for the mirror nucleus ${}_{Z}^{A}$$N$, if the
${}_{N}^{A}$$Z$ is found to be in a toroidal configuration.

An examination of the energies of axially-symmetric toroidal
configurations as a function of $Q_{20}$ in Fig.~\ref{Fig2} reveals
however that they lie on a slope. This indicates that even though the
shell effects cause the density to become toroidal when there is a
quadrupole constraint, the magnitudes of the shell corrections are not
sufficient to stabilize the tori against the bulk tendency to return
to sphere-like geometry.

\subsection{Toroidal configurations with non-collective $I$=$I_z$ rotations }

From our earlier work on the liquid-drop model of a rotating toroidal
nucleus, we expect that an angular momentum about the symmetry axis
will have a stabilizing effect on the toroidal nucleus and the nucleus
may be stabilized against contraction of the torus, when the angular
momentum exceeds a threshold value \cite{Won78}.  We would like to
study whether the toroidal nucleus may indeed be stabilized under a
non-collective rotation about the symmetry axis in the mean-field
theory. Therefore, we take these toroidal configurations obtained for
$I$=0 as the initial configurations and set them in a non-collective
rotation in $Q_{20}$-constrained cranked Skyrme-HF calculations.

In the case of the non-collective rotation about the symmetry axis, the
particle-hole excitations weaken the pairing interaction which can be
approximately neglected for large $I$=$I_z$ in the cranking approach
(see e.g. \cite{Afa99} subsection 3.4).
The search for the final toroidal high-spin isomeric state of a light
nucleus is facilitated by a good starting point at $I$=0, but there is
a great deal of flexibility in choosing this starting point (for example,
by choosing alternatively a chain of alpha particles and omitting the
pairing interaction \cite{Ich12,Ich14a,Ich14b}). The quantitative
measure of the pairing gap for the starting toroidal solution do not
affect sensitively the final toroidal equilibrium (high-spin)
solution we have obtained.

\begin{figure}[htb]
\begin{center}
 \includegraphics[width=0.5\columnwidth]{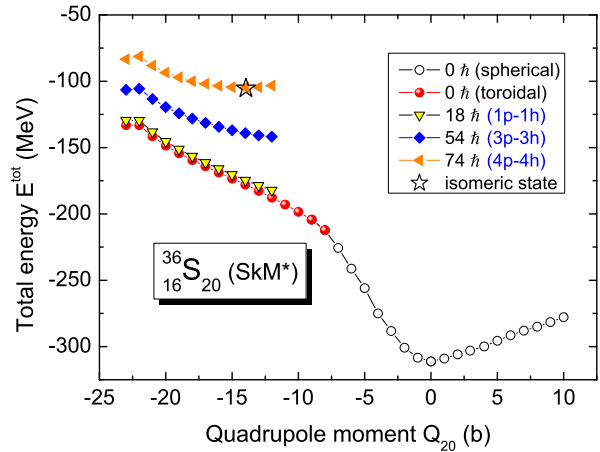}
  \caption{\label{Fig3} (Color online.) The total energy $E^{\rm tot}$ of
    toroidal high-spin states of ${}_{16}^{36}$S as a function of
    $Q_{20}$ for different values of the angular momentum $I$=$I_z$.
    The $n$p-$n$h configurations relative to the $I$=0 configuration
    are also indicated. The location of isomeric toroidal energy
    minimum for $I$=74$\hbar$ is indicated by the star symbol.}
\end{center}
\end{figure}

From the quantum mechanical point of view, the non-collective rotation
around the symmetry axis of an axially-symmetric nucleus corresponds
simply to $n$-particle $n$-hole ($n$p-$n$h) excitations \cite{Boh81,Voi83}.
For this axially-symmetric toroidal even-even nucleus the occupation
of all levels below the Fermi energy leads naturally to the state with
$I$=0. The ($n$p-$n$h) excitations relative to this $I$=0 toroidal
configuration will lead to high-spin toroidal isomeric states with
the total spin given by
\begin{equation}
I=I_z=\sum_{i}(\Omega_{zi}^{part} -\Omega_{zi}^{hole}).
\label{eqI}
\end{equation}
For any given $N$ or $Z$ value and $Q_{20}$ deformation constraint
we need to estimate the Lagrangian multiplier $\hbar\omega$ that
self-consistently leads to the $I$=$I_z$ state in the cranked HF calculation.
To do this we can use Fig.~\ref{Fig1}(b) as a qualitative guide to locate
the ($N,I$) or ($Z,I$) `shells'. Knowing the position of the
shell as a function of $\hbar\omega$, we can appropriately
increase the value of $\hbar\omega$ in the cranked HF model,
to go to the shell in the 'right' (with larger spin) or
decrease $\hbar\omega$ to go to the shell in the 'left' (with smaller
spin). In this way, we can reach all quantized angular momentum $I$
for a specific mass number $A$ as revealed qualitatively by
Fig.~\ref{Fig1}(b). In this connection, it is easy to see that the
shell effects of greater stability can be achieved if the shells for
e.g. ($N,I_n$) and ($Z,I_p$) occur at the same value of $\hbar\omega$,
leading to the ($A, I=I_{n}+I_{p}$) high-spin configuration at some
$Q_{20}$ deformation constraint.

It should be stressed that Fig.~\ref{Fig1}(b) allows us to estimate
the Lagrangian multiplier $\hbar\omega$ only qualitatively and the
'average' (for $A$$\sim$40 and $R/d$=4.5) values of $\hbar\omega$
in Fig.~\ref{Fig1}(b) for the specific shells ($N$ or $Z,I$)
differ from the values presented in Table~\ref{table2} for the
high-spin toroidal isomers with 28$\leq$$A$$\leq$52.

\begin{figure}[htb]
\begin{center}
 \includegraphics[width=0.5\columnwidth]{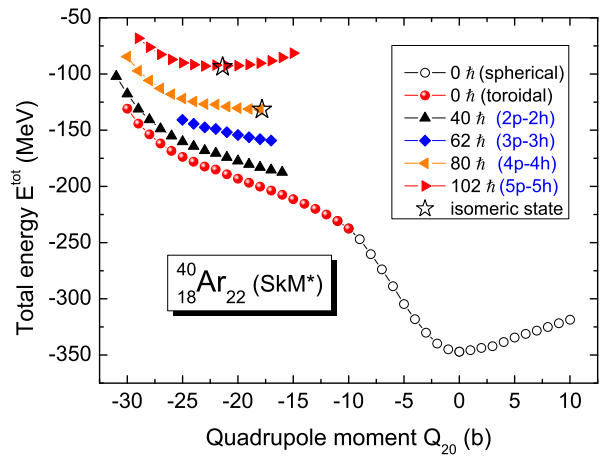}
  \caption{\label{Fig4} (Color online.) The total energy $E^{\rm tot}$ of
    toroidal high-spin states of ${}_{18}^{40}$Ar as a function of
    $Q_{20}$ for different values of the angular momentum $I$=$I_z$.
    The $n$p-$n$h configurations relative to the $I$=0 configuration
    are also indicated. The locations of isomeric toroidal energy
    minima for $I$=80$\hbar$ and $I$=102$\hbar$ are indicated by the
    star symbols.}
\end{center}
\end{figure}

To find the high-spin toroidal isomer configurations for a given mass
number $A$, we use the plots of total energy $E^{\rm tot}$ of the
nucleus with the mass number $A$ as a function of the $Q_{20}$,
calculated in the cranked constrained SkM*-HF model for all quantum
mechanically allowed values of the angular momentum $I$=$I_z$.
At each point ($Q_{20},I$), where the total energy plot reveals
the local energy minimum the quadrupole constraint is removed
and free-convergence is tested to ensure that the toroidal nucleus
is indeed a figure of equilibrium (a high-spin isomeric state).\footnote{
It is worth noting that in these unconstrained and symmetry-unrestricted
cranked Skyrme-HF calculations we do not impose the axial symmetry to the system.}
Figs.~\ref{Fig3} and \ref{Fig4} show the results of this method in
the case of ${}_{16}^{36}$S and ${}_{18}^{40}$Ar, respectively.
We find toroidal high-spin isomers for ${}_{16}^{36}$S($I$=74$\hbar$)
and ${}_{18}^{40}$Ar($I$=82, 102$\hbar$).

The toroidal high-spin isomer ${}_{16}^{36}$S($I$=74$\hbar$) arises
from the shells at $(Z,I)$=(16,33) and $(N,I)$=(20,41) which are
located on the same $\hbar \omega$$\sim$1.8 MeV in Fig.~\ref{Fig1}(b),
corresponding to a 4p-4h excitation of both neutrons and protons
(relative to the $I$=0 configuration). Similarly, the toroidal
high-spin isomer ${}_{18}^{40}$Ar($I$=80$\hbar$) arises from the shells
at $(Z,I)$=(18,36) and $(N,I)$=(22,44) located at $\hbar
\omega$$\sim$1.8 MeV in Fig.~\ref{Fig1}(b), corresponding
to a 4p-4h excitation of both neutrons and protons. On the
other hand, the ${}_{18}^{40}$Ar($I$=102$\hbar$) toroidal
high-spin isomer arises from the shells at $(Z,I)$=(18,46)
and $(N,I)$=(22,56) located at $\hbar \omega$$\sim$2.3 MeV
in Fig.~\ref{Fig1}(b), corresponding to a 5p-5h excitation
of both neutrons and protons.

\subsection{Stability of the toroidal high-spin isomers against nucleon emission}

\begin{figure}[htb]
\begin{center}
  \includegraphics[height=0.6\columnwidth,angle=-90]{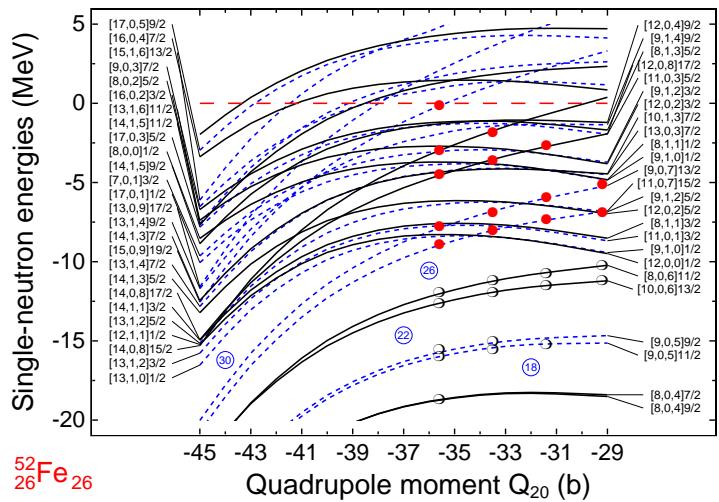}
  \caption{\label{Fig5} (Color online.) The
    neutron-quasiparticle energies as a function of the quadrupole
    moment $Q_{20}$ obtained in the SkM*-HF+BCS model for toroidal $^{52}$Fe with
    $I$=0. They are labeled by $[Nn_{z}\Lambda]\Omega$, with even
    parity levels as solid lines, and odd parity levels as dashed
    lines. Starting from the $I$=0 configuration, the 2p-2h, 3p-3h, 4p-4h, and 5p-5h
    excitations shown in the plot (with holes as open circles and
    particles in solid circular points) lead to non-collective rotations with a
    total $I_z$=26, 40, 52, and 66$\hbar$, respectively.
    The horizontal (red) dashed line represents energy $E=0$.}
\end{center}
\end{figure}

\begin{figure}[htb]
\begin{center}
  \includegraphics[height=0.6\columnwidth,angle=-90]{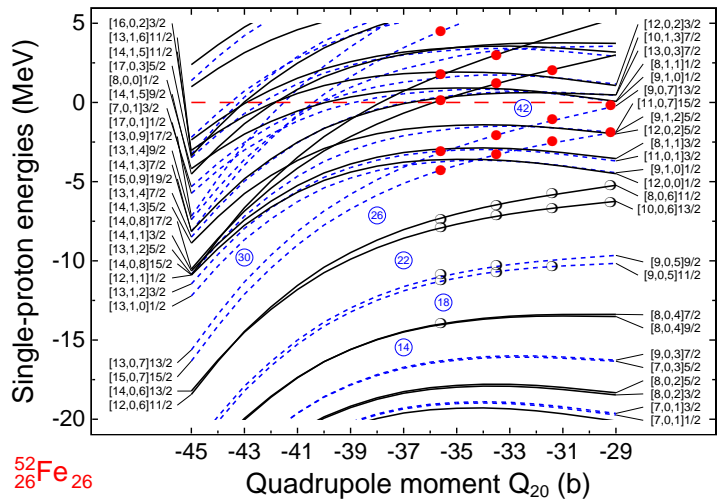}
  \caption{\label{Fig6} (Color online.) The same as in Fig.~\ref{Fig5}, but for
  the proton-quasiparticle energies.}
\end{center}
\end{figure}

In Figs.~\ref{Fig5} and \ref{Fig6}, the neutron- and proton-quasiparticle energies
obtained in the SkM*-HF+BCS model\footnote{In this subsection we examine
the single-particle wave functions using the constraint HF+BCS model instead of
constraint HFB approach, but all presented results are valid for both models.
A three-dimensional Cartesian deformed HO basis used in HF+BCS model
is exactly the same as in HFB model, and consists of the lowest 1140 states that
originate from the $N_0$=26 oscillator shells.}
for toroidal $^{52}$Fe with $I$=0 (no rotation) are presented as a function of
the quadrupole moment $Q_{20}$. Each of these states is labeled by asymptotic
quantum numbers $[Nn_{z}\Lambda]\Omega$. One can see that all states lying below
the Fermi level have the quantum numbers $n_{z}=0$ and $\Lambda=0,1,2,\ldots,6$.
To illustrate $n$p-$n$h excitations observed in toroidal $^{52}$Fe we used
solid circular (red) points to mark the high-$\Lambda$ particle states
excited from lower-$\Lambda$ hole states marked by open circular points.
Each of the $n$p-$n$h excitation configurations is shown at the quadrupole
deformation of the toroidal high-spin isomer of $^{52}$Fe (see Table~\ref{table2}).
It should be mentioned that all occupied excited (particle) states possess
the same quantum number $n_{z}=0$ as the hole states, and have
($\Lambda,\Omega$)=(7,15/2), (7,13/2), (8,17/2), (8,15/2), (9,19/2).
While the neutron quasiparticle energies of occupied excited states are
all negative (Fig.~\ref{Fig5}), the proton quasiparticle energies
of the occupied high-$\Lambda>$7 states are positive (Fig.~\ref{Fig6}),
raising questions whether toroidal high-spin isomers obtained by
occupying these high-$\Lambda$ proton states are stable against
nucleon emission.

\begin{figure}[htb]
\begin{center}
  \includegraphics[width=0.5\columnwidth]{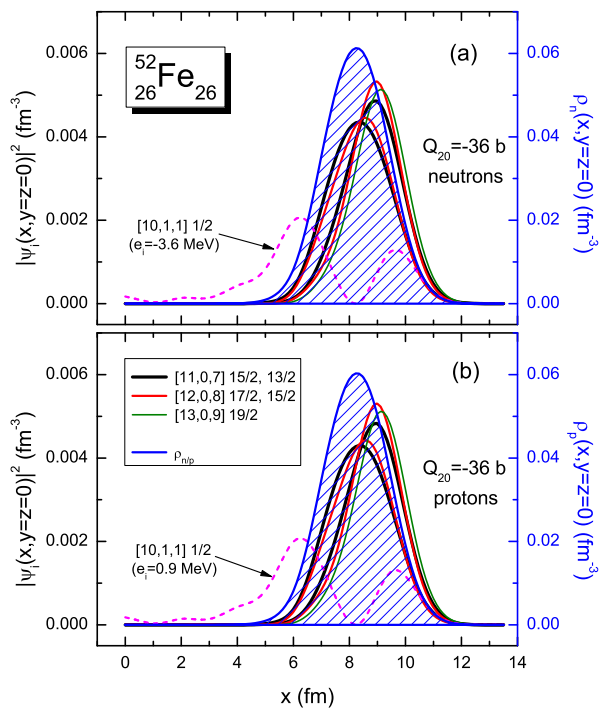}
  \caption{\label{Fig7} (Color online.) (a) The modulus squared of the
    neutron wave functions $[N,n_{z}$=0,$\Lambda$=7,8,9$]\Omega$,
    representing the particle states in the $n$p-$n$h excitations of
    $^{52}$Fe (see Figs.~\ref{Fig5} and \ref{Fig6}). The state wave
    functions and the density distributions $\rho_{n}$ (the hatched area)
    were calculated in the SkM*-HF+BCS model with a constraint on the
    quadrupole moment, $Q_{20}$=-36 b, which is close to the deformation
    of the toroidal $^{52}$Fe(132$\hbar$) isomer. On the same plot is shown,
    as an example, the modulus squared of the state wave function
    $[10,1,1]1/2$ which is not used in the construction of the
    toroidal configurations.
    Panel (b) the same as in the panel (a), but for the proton wave
    functions and proton density distribution $\rho_{p}$.}
\end{center}
\end{figure}

\begin{figure}[htb]
\begin{center}
  \includegraphics[width=0.5\columnwidth]{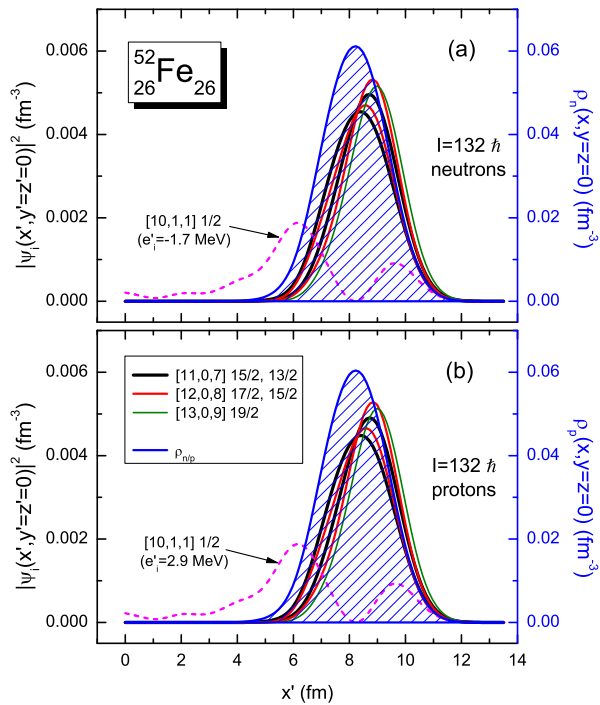}
  \caption{\label{Fig8} (Color online.) The same as in Fig.~\ref{Fig7},
  but the wave functions and the neutron/proton density distributions
  were calculated in the cranked SkM*-HF model for the equilibrium
  configuration of the isomeric toroidal state $^{52}$Fe(132$\hbar$).}
\end{center}
\end{figure}

To address this issue, we would like to focus on 5p-5h
excitation at $Q_{20}=-35.6$~b. For $N$ or $Z=$26 we can
estimate using Table~\ref{table1} that $\Lambda_{max}=6$
and $I=10\Lambda_{max}+6=66 \hbar$ for neutrons and protons,
so the total angular momentum along the symmetry
axis is $I$=132$\hbar$ in this case. Fig.~\ref{Fig7}
displays the modulus squared of the neutron (panel (a))
and proton (panel (b)) wave functions of $^{52}$Fe
with $n_{z}$=0 and $\Lambda$=7,8,9, calculated along
$x$-direction in the SkM*-HF+BCS model with constraint
$Q_{20}=$$-$36~b. In the same plot we display also
the neutron/proton density distributions $\rho_{n/p}$ presented
as hatched areas. The same wave functions as in Fig.~\ref{Fig7}
and the neutron/proton density distributions calculated in
the cranked SkM*-HF model in the intrinsic frame ($x',y',z'$)
for the toroidal high-spin isomer $^{52}$Fe($I$=132$\hbar$)
are shown in Fig.~\ref{Fig8}.

We find that these $n_{z}$=0 and $\Lambda$=7,8,9 wave functions
for both $I$=0 (Fig.~\ref{Fig7}) and $I$=132$\hbar$ (Fig.~\ref{Fig8})
do not exhibit the unbound characteristics of leakage and
oscillation beyond the single-particle potential. They are well
localized in the toroidal region of the attractive mean-field
potential and they have Gaussian shapes very similar to those wave
functions of lower-lying bound single-particle states.
In contrast, the wave functions of the [10,1,1]1/2 state in
Figs.~\ref{Fig7} and \ref{Fig8} that are not used for the construction
of the toroidal isomer - extend outside the single-particle potential.
We also tested our model for a well-known unbound $3s_{1/2}$ state
in the ground state of $^{32}$Fe (at $Q_{20}=1.6$~b) and we found that
the $3s_{1/2}$ single-particle energy is positive and the spatial
wave function leaks to the region of large $r$ with an oscillating
amplitude extending beyond the single-particle potential.
This indicates that deformed HO basis used in our study
is sufficient to properly describe the scattering (unbound) states
and our model allows us to discern different properties of the
scattering and bound states.

There may be two possible reasons why there are no apparent wave
function leakage and oscillations at large $r$ for these $n_{z}$=0
and $\Lambda$=7,8,9 states used for the construction of the toroidal
high-spin isomer: (i) the 'confinement' of these single-particle states
with exponentially decaying probability to reach $r\to \infty$
beyond the single-particle potential or (ii) the presence of large
centrifugal and additional proton Coulomb barriers, allowing only a
small penetration probability for tunneling to $r\to \infty$.

To study further which of the above two possibilities
pertains to the wave functions of the high-$\Lambda$ states
for our toroidal isomers in question, we consider the single-particle
equation
\begin{equation}
(-\frac{\hbar^{2}}{2m}\triangle + V(r,z))\Psi_{E}(\boldsymbol{r})=
E\Psi_{E}(\boldsymbol{r}),
\label{eq1}
\end{equation}
where $V(r,z)= U(r) + W(z)$ is an axially symmetric non-central potential
with a minimum at $V(R,0)$. We assume that
$\lim_{r\to \infty}U(r)=0$ and $\lim_{|z|\to \infty}W(z)=0$.
If the potential is analytic near its minimum, we can make a Taylor
expansion about the point ($r, z$)=($R,0$)
\begin{eqnarray}
V(r,z) &=& U(R) + W(0) + \left.\frac{dU}{dr}\right|_{r=R}(r -R) +
\left.\frac{dW}{dz}\right|_{z=0}z \nonumber\\
&&+\left.\frac{1}{2!} \frac{1}{r}\frac{d}{dr}\left(r \frac{dU}{dr}\right)
\right|_{r=R} (r -R)^{2}+
\left.\frac{1}{2!} \frac{d^{2}W}{dz^{2}}\right|_{z=0}z^{2} + \cdots.
\label{eq11}
\end{eqnarray}
At the minimum the first-order derivatives vanish and we can set constants
to zero, $U(R)=W(0)=0$. If we now shift a radial coordinate $r$ to $r+R$
Eq. (\ref{eq11}) reads
\begin{eqnarray}
V(r,z) &=& \left.\frac{1}{2!} \frac{d^{2}U}{dr^{2}}\right|_{r=R} r^{2}+
\left.\frac{1}{2!} \frac{d^{2}W}{dz^{2}}\right|_{z=0}z^{2} + \cdots \\
&\approx& U(r) + \frac{1}{2}m\omega^{2}_{z}z^{2},
\label{eq12}
\end{eqnarray}
where
\begin{equation}
\omega^{2}_{z}= \left.\frac{1}{m}\frac{d^{2}W}{dz^{2}}\right|_{z=0}.
\label{eq12.1}
\end{equation}

The high-spin toroidal isomers under consideration have been
constructed with single-particle states whose wave functions
in the $z$-direction reside in the lowest state, with a zero
number of nodes, $n_z$=0.
Using the approximation (\ref{eq12}), we can express energy
associated with a $z$-degree of freedom of the $n_z$=0 bound
states by the zero-point energy $E_{0z}= \frac{1}{2}\hbar\omega_{z}$.

Writing $\Psi=P(r) \frac{1}{\sqrt{2\pi}} e^{i\Lambda\varphi} Z_{0}(z)$,
where $Z_{0}(z)=\left(\frac{a}{\pi}\right)^{1/4} e^{-az^{2}/2}$
with $a=m\omega_{z}/\hbar$, and using the method of separation
of variables we receive the radial equation
\begin{equation}
\left[\frac{d^{2}}{dr^{2}} + \frac{1}{r}\frac{d}{dr} + \frac{2m}{\hbar^2}
(E_{\bot}(i)- U(r)) - \frac{\Lambda^2}{r^2} \right]P_{i}(r)=0,
\label{eq14}
\end{equation}
where $\Lambda$ can have any integer value and the energy
of the $r$-direction (transverse) motion, $E_{\bot}(i)$, is equal to the
difference between a total singe-particle energy, $E(i)$,
and the zero-point energy for the motion in the $z$-direction:
\begin{equation}
E_{\bot}(i)=E(i)-\frac{1}{2}\hbar\omega_{z}.
\label{eqe}
\end{equation}
We provide the solution to Eq. (\ref{eq14}) in Appendix A,
where we use a finite square potential well to describe the
radial potential $U(r)$.

The bound states of the nucleon occur when the energy
available for motion in the $r$-direction is $E_{\bot}(i)$$<$0.
We have carried out an analysis to check the signs of the
$E_{\bot}(i)$ values for those occupied single-particle
states in the toroidal $^{52}$Fe($I$=132$\hbar$) isomer
for which the signs of total single-particle energies $E(i)$
are positive. The zero-point energy $\frac{1}{2}\hbar\omega_{z}$
for the single-particle states can be extracted from the plots
of their modulus squared wave functions $|\psi_{i}(r=R,z)|^2$
plotted in the $z$-direction. For example, the zero point energy
for the topmost occupied state, [N,0,9]19/2, in the toroidal
high-spin isomer $^{52}$Fe($I$=132$\hbar$) is equal 8.0 MeV
for neutrons and protons.
From these $E_{0z}=\frac{1}{2}\hbar\omega_{z}$ and Eq. (\ref{eqe}),
the single-particle energies for radial motion $E_{\bot}(i)$
can be calculated.
The self-consistent calculations in the toroidal isomer
$^{52}$Fe($I$=132$\hbar$) reveal that $E_{\bot}(i)$ for the
$n_z$=0 and $\Lambda$=7,8,9 states are negative, leading to
exponentially decaying single-particle radial wave functions,
$P_{i}(r)$, for large values of $r$.
This indicates that wave functions of these occupied
single-particle states in toroidal $^{52}$Fe($I$=132$\hbar$)
isomer are not only localized but also square integrable.
However, it should be stressed that any $n$p-$n$h particle state
with $E(i)>\frac{1}{2}\hbar\omega_z$ will not hold a bound state.
Thus, one expects an $I_z$-window for which bound states are
possible.

It is worth noting that what we have observed are the states with
the positive total energy $E(i)$ whose wave functions
are localized in the $z$- and $r$-directions and are square integrable.
These states seem analogous to the bound states in the continuum (BIC)
first suggested by von Neumann and Wigner in 1929 \cite{Neu29}
(see also \cite{Sti75}) and recently examined by many workers
in various quantum and optical systems (see e.g. \cite{Pro13} and
references cited therein).

\section{Properties of $N$$\ne$$Z$ toroidal high-spin isomers }

\begin{figure}[htb]
\begin{center}
  \includegraphics[width=0.5\columnwidth]{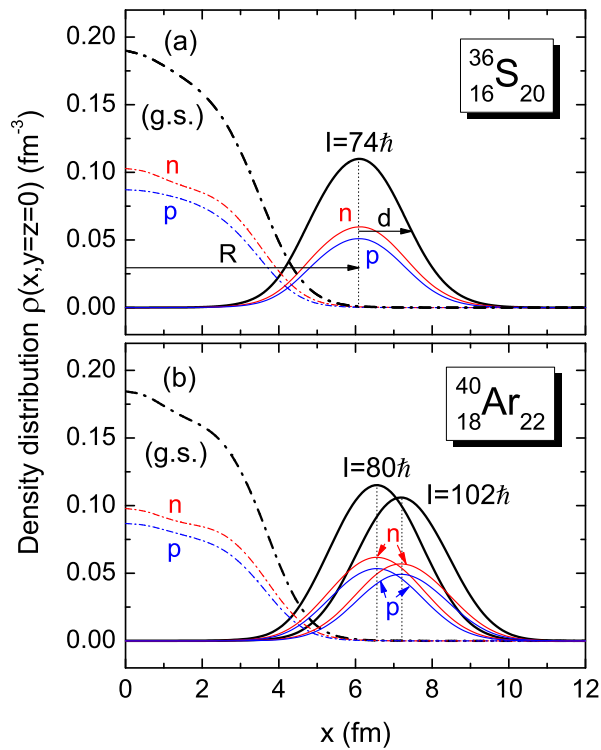}
  \caption{\label{Fig9} (Color online.) (a) The total density
    distribution of the isomeric toroidal state of ${}_{16}^{36}$S
    with $I$=74$\hbar$ as a cut in the radial direction $x$. We show
    the major radius $R$ and the minor radius $d$ of toroidal density
    distribution.  The dash-dot curve shows the total
    density distribution in the ground state (g.s.) of
    ${}_{16}^{36}$S.  Lines indicated by n/p represent the
    neutron/proton density distributions, respectively.  (b) Total
    density distributions of the isomeric toroidal states of
    ${}_{18}^{40}$Ar with $I$= 80 and 102$\hbar$ as a cut in the
    radial direction $x$, all symbols are the same as in (a).}
\end{center}
\end{figure}

After we have located the toroidal high-spin isomers at their energy
minima, we can evaluate their properties.
We plot in Fig.~\ref{Fig9}(a) and (b) the density distributions
of the toroidal isomers ${}_{16}^{36}$S($I$=74$\hbar$) (presented in Fig.~\ref{Fig3})
and ${}_{18}^{40}$Ar($I$=80, 102$\hbar$)
(Fig.~\ref{Fig4}), as a cut in the radial direction $x$.
For comparison, we also show the total density distribution of
${}_{16}^{36}$S and ${}_{18}^{40}$Ar in their ground state (g.s.)
(dash-dot curve) which have a maximum $\rho_{max}$ value distinctly
larger than the $\rho_{max}$ of the toroidal high-spin isomers.
This is a general property similar to those of other $N$=$Z$ toroidal
high-spin isomers~\cite{Sta14}.
In the case of ${}_{18}^{40}$Ar nucleus when the aligned angular momentum $I$
increases from $I$=80$\hbar$ to $I$=102$\hbar$, the maximum toroidal
density $\rho_{max}$ decreases from 0.116 to 0.107 fm$^{-3}$ and the
major radius $R$ increases from 6.56 to 7.21 fm. Only the minor radius
$d$, defined as a half width at half maximum (HWHM) of the toroidal
distribution, stays constant at $d\approx$1.37 fm.
In addition to the total density distributions we compare in
Fig.~\ref{Fig5} the neutron (n) and proton (p) density distributions
in the toroidal isomeric states as well as in g.s. of ${}_{16}^{36}$S
and ${}_{18}^{40}$Ar.

\begin{table}[htb]
%\begin{center}
\caption{\label{table2} Properties of the high-spin toroidal isomers
  at their local energy minima in 28$\le$$A$$\le$52.
  For nuclei with N=Z all values are taken from Ref.~\cite{Sta14}.
  The angular momentum $I$ is a step-wise function for a range of
  $\hbar\omega$, for which we give one of the points in the range.
}
\vspace*{0.3cm}
\resizebox{\columnwidth}{!} {
\begin{tabular}{lrcd{3}ccccc}
\toprule
          & $I/\hbar$ & $Q_{20}$ & \multicolumn{1}{c}{$\hbar\omega$} &
$E^*$     & $R$       &   $d$    & $R/d$ & $\rho_{\rm max}$ \\
          &           & (b)      &  \multicolumn{1}{c}{(MeV)}        &
(MeV)     &~(fm)~     &~(fm)~&       & (fm$^{-3}$)  \\ \midrule
${}_{14}^{28}$Si &        44 &  -5.86   &  2.8          & 143.18 & 4.33  & 1.45 & 2.99  & 0.119 \\
${}_{16}^{32}$S  &        48 &  -8.22   &  1.9          & 153.87 & 4.87  & 1.42 & 3.43  & 0.122 \\
                 &        66 & -10.51   &  2.2          & 193.35 & 5.57  & 1.40 & 3.98  & 0.108 \\
${}_{16}^{36}$S  &        74 & -13.95   &  1.85         & 205.87 & 6.08  & 1.39 & 4.37  & 0.112 \\
${}_{18}^{36}$Ar &        56 & -11.31   &  1.7          & 168.03 & 5.44  & 1.40 & 3.88  & 0.125 \\
                 &        72 & -13.73   &  1.85         & 198.63 & 6.04  & 1.39 & 4.34  & 0.113 \\
                 &        92 & -16.78   &  2.0          & 238.56 & 6.73  & 1.37 & 4.91  & 0.103 \\
${}_{18}^{40}$Ar &        80 & -17.83   &  1.65         & 215.49 & 6.56  & 1.38 & 4.75  & 0.116 \\
                 &       102 & -21.37   &  1.85         & 253.42 & 7.21  & 1.37 & 5.26  & 0.107 \\
${}_{20}^{40}$Ca &        60 & -14.96   &  1.5          & 178.36 & 5.97  & 1.40 & 4.26  & 0.126 \\
                 &        82 & -17.61   &  1.9          & 214.23 & 6.51  & 1.39 & 4.68  & 0.117 \\
${}_{22}^{44}$Ti &        68 & -19.57   &  1.2          & 195.46 & 6.55  & 1.39 & 4.71  & 0.128 \\
                 &        88 & -22.27   &  1.4          & 223.09 & 7.01  & 1.38 & 5.08  & 0.120 \\
                 &       112 & -25.76   &  1.6          & 260.24 & 7.56  & 1.37 & 5.52  & 0.113 \\
${}_{24}^{48}$Cr &        72 & -25.08   &  1.2          & 207.12 & 7.12  & 1.38 & 5.16  & 0.128 \\
                 &        98 & -28.00   &  1.4          & 239.26 & 7.54  & 1.37 & 5.50  & 0.122 \\
                 &       120 & -30.55   &  1.43         & 271.02 & 7.90  & 1.36 & 5.81  & 0.118 \\
${}_{26}^{52}$Fe &        52 & -29.24   &  0.8          & 202.86 & 7.39  & 1.38 & 5.35  & 0.134 \\
                 &        80 & -31.43   &  0.95         & 227.54 & 7.68  & 1.38 & 5.56  & 0.130 \\
                 &       104 & -33.54   &  1.3          & 252.65 & 7.94  & 1.37 & 5.79  & 0.126 \\
                 &       132 & -35.62   &  1.5          & 288.91 & 8.20  & 1.36 & 6.03  & 0.123 \\
\bottomrule
\end{tabular}
}
%\end{center}
\end{table}

\begin{figure}[htb]
\begin{center}
  \includegraphics[width=0.5\columnwidth]{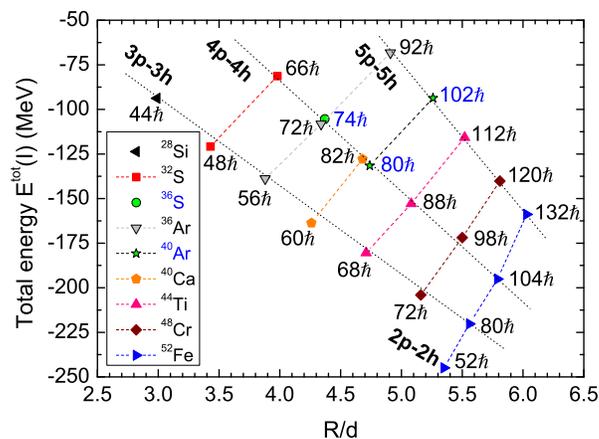}
  \caption{\label{Fig10} (Color online.) The total energies of the
    isomeric toroidal states of ${}_{14}^{28}$Si, ${}_{16}^{32}$S,
    ${}_{16}^{36}$S, ${}_{18}^{36}$Ar, ${}_{18}^{40}$Ar,
    ${}_{20}^{40}$Ca, ${}_{22}^{44}$Ti, ${}_{24}^{48}$Cr,
    and ${}_{26}^{52}$Fe and their associated $I$ values (Table~\ref{table1}),
    as a function of $R/d$. The $n$p-$n$h configurations relative to the $I$=0
    configuration are also indicated.}
\end{center}
\end{figure}

With the additional information on the $N$$\ne$$Z$ isomers, we collect
the properties of all known 21 toroidal high-spin isomers in
Table~\ref{table2} and Fig.~\ref{Fig10}.
In Table~\ref{table2}, we list the quantized angular momentum $I$ of the
toroidal isomers, its corresponding quadrupole moment $Q_{20}$,
the excitation energy $E^*$ of the isomer relative to the ground state
configuration, the major radius $R$ of the toroid, the minor radius $d$
of the toroid, and the maximum density of the isomer $\rho_{\max}$.
The angular momentum $I$ is a step-wise function of the cranking frequency
$\hbar \omega$ \cite{Ring80}, and there is a range of the cranking frequency
(energy) $\hbar \omega$ at which the toroidal high-spin isomer is located.
We list the value $\hbar \omega$ of a point within the range in Table~\ref{table2}.
These quantities provide a wealth of information about the high-spin
toroidal isomers from which properties on the nuclear fluid in the
exotic toroidal shape may be extracted.

It is useful to classify the isomers according to their $n$p-$n$h
attributes relative to their corresponding $I$=0 configurations.
One finds that the all $n$p-$n$h families follow regular well-behaved pattern
as shown in Fig.~\ref{Fig10}, where we plot the total energy of the toroidal
isomers as a function of the toroidal aspect ratio $R/d$.
The corresponding angular momentum associated with each isomer is also listed.
It is important to notice that $N$$\ne$$Z$ nuclei fit very well to the same pattern
as $N$=$Z$ nuclei, indicating a smooth behavior for all even-even toroidal
high-spin isomers.

Table~\ref{table2} further reveals that in each $n$p-$n$h family,
the angular momentum and the aspect ratio $R/d$
increase approximately linearly with the mass number while
the minor radius $d$ remains essentially unchanged.
One can use Table~\ref{table2} and Fig.~\ref{Fig10}
to extrapolate the properties of toroidal high-spin isomers
in the higher mass region.

\section{Conclusions and discussion}

Nuclei under non-collective rotation with a large angular momentum
about the symmetry axis above a threshold can assume a toroidal shape.
We have developed a systematic method of cranked SHF
approach by which the high-spin toroidal isomer states can be
theoretically located. The application of the method to study nuclei
in all regions of the periodical table will enhance our knowledge of
the multi-faceted nature of the nuclear phenomenon, allow us to
explore into the possibility of toroidal isomers as a source of
energy, and permit the study of nuclear fluid in extreme geometry and
angular momenta.

Our investigation into the region of nuclei with $N$$\ne$$Z$ indicates
that just as the $N$=$Z$ nuclei, toroidal high-spin nuclei are also
commonly present. In the present work, we have located the toroidal
${}_{16}^{36}$S($I$=74$\hbar$) and ${}_{18}^{40}$Ar($I$=80, 102$\hbar$)
isomers.  As indicated in Fig.~\ref{Fig2}, many other nuclei with
other neutron and proton numbers also have axial-symmetric toroidal
density distributions under a large-magnitude negative quadrupole
moment constraint. We expect that when these nuclei are set to
undergo non-collective rotation beyond a threshold, toroidal energy
minima and additional toroidal high-spin isomers will be present.
Furthermore, because of the approximate symmetry of $N$ and $Z$,
toroidal configurations are also expected for the mirror nucleus
${}_{Z}^{A}$$N$, if the ${}_{N}^{A}$$Z$ is found to be in a toroidal
configuration.
The results obtained here and in Ref.~\cite{Sta14} provide a wealth of information
to allow the extrapolation in future search of other toroidal nuclei.
Extrapolation from Table~\ref{table2} and Fig.~\ref{Fig10} would predict
possible occurrence of toroidal high-spin isomers in the mass region
of $A$$\sim$60.

Returning to the questions we have posed in the beginning, the
occurrence of these $N$$\ne$$Z$ toroidal high-spin isomers shows that
it is not necessary to be $\alpha$-particle-type nuclei to be a
toroidal high-spin isomer.  The conditions for the occurrence of the
toroidal high-spin isomer consist of (i) the occurrence of favorable
shell-model configuration such as those indicated by the
single-particle diagram similar to Figs.~\ref{Fig1}(a) and (b), and
(ii) the quantized angular momentum value exceeding a threshold value.
There is a third condition on the maximum limit of $I$=$I_z$ and its
relation to sausage instabilities that has been examined only for a
few cases at the present moment and has not yet received sufficient
attention to render a definitive conclusion.

\section*{Acknowledgements}
This work was supported in part by the Division of Nuclear
Physics, U.S. Department of Energy, Contract No. DE-AC05-00OR22725.

\section*{Appendix A}

In this appendix we provide the solution to the radial equation
\begin{equation}
\left[\frac{d^{2}}{dr^{2}} + \frac{1}{r}\frac{d}{dr} + \frac{2m}{\hbar^2}
(E_{\bot}(i)- U(r)) - \frac{\Lambda^2}{r^2} \right]P_{i}(r)=0,
\label{eq14a}
\end{equation}
where $E_{\bot}(i)=E(i)-\frac{1}{2}\hbar\omega_{z}$,
and $\Lambda$ can have any integer value.

To solve radial equation (\ref{eq14a}) we use a square well
approximation for the radial potential $U(r)$:
\begin{equation}
U(r)=\left\{ \begin{array}{ll}
U_{0} (<0) & \textrm{for } R-d \leq r \leq R+d \\
0          & \textrm{for } R-d > r > R+d, \end{array}\right.
\label{eq15}
\end{equation}
where $R$ and $d$ are the major- and minor-radius of torus,
respectively. Then, the radial function, for bound states
with $-|U_{0}|< E_{\bot}(i)< 0$, obeys
\begin{equation}
r^{2}\frac{d^{2}P_{i}}{dr^{2}} + r\frac{dP_{i}}{dr} + \left(k^{2}_{0z}r^{2}
- \Lambda^2\right)P_{i}=0,
\label{eq16}
\end{equation}
where the radial wavenumber, $k$, is defined by
\begin{equation}
k^{2}_{0z}=\left\{ \begin{array}{ll}
\frac{2m}{\hbar^2}(E_{\bot}(i)+ |U_{0}|)\equiv \alpha^{2}
& \textrm{for } R-d \leq r \leq R+d \\
\frac{2m}{\hbar^2}E_{\bot}(i)\equiv -\beta^{2} &
\textrm{for } R-d > r > R+d.
\end{array}\right.
\label{eq16.1}
\end{equation}

If we change variable $r$ to $\alpha r$ in the region
$R-d \leq r \leq R+d$ the radial function $P_i$ satisfies
the Bessel's ODE. Similarly, if we change variable $r$
to $i\beta r$ in $R-d > r > R+d$ the radial function $P_i$
satisfies the modified Bessel's ODE. A general solution
for the radial function takes a form
\begin{equation}
P_{i}=\left\{ \begin{array}{ll}
A'I_{|\Lambda|}(\beta r) + B'K_{|\Lambda|}(\beta r) &
\textrm{for }   r < R-d \\
AJ_{|\Lambda|}(\alpha r) + BY_{|\Lambda|}(\alpha r) &
\textrm{for } R-d \leq r \leq R+d \\
A''I_{|\Lambda|}(\beta r) + B''K_{|\Lambda|}(\beta r) &
\textrm{for } r > R+d, \end{array}\right.
\label{eq17}
\end{equation}
where we use the standard Bessel function of the first kind,
$J_{\Lambda}$, and Bessel function of the second kind,
$Y_{\Lambda}$, as well as the modified Bessel function
of the first kind, $I_{\Lambda}$, and the modified Bessel
function of the second kind (the modified Hankel function),
$K_{\Lambda}$. Since for integer order $\Lambda$, $J_{\pm\Lambda}$,
$Y_{\pm\Lambda}$, $I_{\pm\Lambda}$, and $K_{\pm\Lambda}$
are not linearly independent, we use the absolute value of
$\Lambda$ in Eq. (\ref{eq17}).

The radial function must be finite everywhere including at the origin,
therefore the constant $B'$ has to be equal zero, because the modified
Bessel function of the second kind, $K_{|\Lambda|}$, is singular at the origin.
Also $A''=0$, due to divergent behavior of $I_{|\Lambda|}$ at infinity.
Finally, the solution of radial equation (\ref{eq16}) for an energy in
the interval $-|U_{0}|< E_{\bot}(i)< 0$ reads
\begin{equation}
P_{i}=\left\{ \begin{array}{ll}
A'I_{|\Lambda|}(\beta r) & \textrm{for }   r < R-d \\
AJ_{|\Lambda|}(\alpha r) + BY_{|\Lambda|}(\alpha r) &
\textrm{for } R-d \leq r \leq R+d \\
B''K_{|\Lambda|}(\beta r) & \textrm{for } r > R+d.
\end{array}\right.
\label{eq18}
\end{equation}

The equations that permit us to calculate the eigenvalues are
the continuity condition of the logarithmic derivative of
$P_{i}$ at $r=R-d$ and $r=R+d$:
\begin{eqnarray}
\left.\frac{I'_{|\Lambda|}(\beta r)}{I_{|\Lambda|}(\beta r)}\right|_{r=R-d}
=\left.\frac{AJ'_{|\Lambda|}(\alpha r)+ BY'_{|\Lambda|}(\alpha r)}
{AJ_{|\Lambda|}(\alpha r)+ BY_{|\Lambda|}(\alpha r)}\right|_{r=R-d},\\
\left.\frac{K'_{|\Lambda|}(\beta r)}{K_{|\Lambda|}(\beta r)}\right|_{r=R+d}
=\left.\frac{AJ'_{|\Lambda|}(\alpha r)+ BY'_{|\Lambda|}(\alpha r)}
{AJ_{|\Lambda|}(\alpha r)+ BY_{|\Lambda|}(\alpha r)}\right|_{r=R+d},
\label{eq19}
\end{eqnarray}
where we shall adopt the notation that the prime is derivative
with respect to $r$. Using formulas for derivatives of the Bessel
functions and the modified Bessel functions the continuity condition
equations take the form
\begin{eqnarray}
\beta\left.\frac{I_{|\Lambda|+1}(\beta r)}{I_{|\Lambda|}(\beta r)}\right|_{r=R-d}&=&
-\alpha\left.\frac{\mathcal{Z}_{|\Lambda|+1}(\alpha r)}
{\mathcal{Z}_{|\Lambda|}(\alpha r)}\right|_{r=R-d},\label{eq19.3a}\\
\beta\left.\frac{K_{|\Lambda| +1}(\beta r)}{K_{|\Lambda|}(\beta r)}\right|_{r=R+d}&=&
\alpha\left.\frac{\mathcal{Z}_{|\Lambda|+1}(\alpha r)}
{\mathcal{Z}_{|\Lambda|}(\alpha r)}\right|_{r=R+d},
\label{eq19.3b}
\end{eqnarray}
where we introduce shorthand notation (a cylinder function $\mathcal{Z}$)
\begin{equation}
\mathcal{Z}_{|\Lambda|}(\alpha r)= AJ_{|\Lambda|}(\alpha r)+ BY_{|\Lambda|}(\alpha r).
\end{equation}
Eliminating constants $A$ and $B$ from Eqs. (\ref{eq19.3a}) and (\ref{eq19.3b})
we obtain the eigenvalue equation
\begin{eqnarray}
&&\alpha^{2}\; I_{|\Lambda|}(\beta(R-d)) K_{|\Lambda|}(\beta(R+d))\nonumber\\
&&\left[J_{|\Lambda|+1}(\alpha(R+d)) Y_{|\Lambda|+1}(\alpha(R-d))-
      J_{|\Lambda|+1}(\alpha(R-d)) Y_{|\Lambda|+1}(\alpha(R+d))\right]\nonumber\\
&+&\beta^{2}\; I_{|\Lambda|+1}(\beta(R-d)) K_{|\Lambda|+1}(\beta(R+d))\nonumber\\
&&\left[J_{|\Lambda|}(\alpha(R-d)) Y_{|\Lambda|}(\alpha(R+d))-
      J_{|\Lambda|}(\alpha(R+d)) Y_{|\Lambda|}(\alpha(R-d))\right]\nonumber\\
&+&\alpha\beta\; I_{|\Lambda|+1}(\beta(R-d)) K_{|\Lambda|}(\beta(R+d))\nonumber\\
&&\left[J_{|\Lambda|+1}(\alpha(R+d)) Y_{|\Lambda|}(\alpha(R-d))-
      J_{|\Lambda|}(\alpha(R-d)) Y_{|\Lambda|+1}(\alpha(R+d))\right]\nonumber\\
&+&\alpha\beta\; I_{|\Lambda|}(\beta(R-d)) K_{|\Lambda|+1}(\beta(R+d))\nonumber\\
&&\left[J_{|\Lambda|+1}(\alpha(R-d)) Y_{|\Lambda|}(\alpha(R+d))-
      J_{|\Lambda|}(\alpha(R+d)) Y_{|\Lambda|+1}(\alpha(R-d))\right]\nonumber\\
&=&0,
\label{eq19.4}
\end{eqnarray}
where $\alpha=\sqrt{\frac{2m}{\hbar^2}(|U_{0}|-|E_{\bot}(i)|)}$,
$\beta=\sqrt{\frac{2m}{\hbar^2}|E_{\bot}(i)|}\neq 0$.

There are no solutions to Eq. (\ref{eq19.4}) if $E_{\bot}(i)< -|U_{0}|$.
In the energy range $-|U_{0}|< E_{\bot}(i) < 0$ there is the discrete spectrum.
Since $E_{\bot}(i)= 0$ is a continuous spectrum limit, we see that
in terms of $E(i)= E_{\bot}(i)+ \frac{1}{2}\hbar\omega_{z}$
the continuum limit is equal to the zero point energy of
the $z$-direction motion $E(i)=\frac{1}{2}\hbar\omega_{z}$.
The continuous spectrum exists for $E>\frac{1}{2}\hbar\omega_{z}$
($E_{\bot}> 0$) and the radial wave function takes the form
\begin{equation}
P=\left\{ \begin{array}{ll}
A'J_{|\Lambda|}(\beta r) & \textrm{for }   r < R-d \\
AJ_{|\Lambda|}(\alpha r) + BY_{|\Lambda|}(\alpha r) &
\textrm{for } R-d \leq r \leq R+d \\
A''J_{|\Lambda|}(\beta r) + B''Y_{|\Lambda|}(\beta r) &
\textrm{for } r > R+d, \end{array}\right.
\label{eq19.6}
\end{equation}
with $\alpha^{2}=\frac{2m}{\hbar^2}(E_{\bot}+ |U_{0}|)$ and
$\beta^{2}=\frac{2m}{\hbar^2}E_{\bot}$.

We would like to point out that similar quantum systems supporting
an old idea of a so-called bound state in the continuum (BIC),
suggested by von Neumann and Wigner \cite{Neu29},were discussed
by Robnik \cite{Rob86} and recently in Ref.~\cite{Pro13}.
\\

\end{document}